\documentclass[pra,preprintnumbers,floatfix,letterpaper,superscriptaddress,footinbib,twocolumn,10pt]{revtex4-1}
\usepackage[utf8x]{inputenc}

\usepackage{amsmath,amssymb,amsthm}

\usepackage{bm}       
\usepackage{graphicx} 
\usepackage{epsfig}   
\usepackage[usenames]{color}
\usepackage{tabularx}

\begin{document}

\title{High-rate field demonstration of large-alphabet quantum key distribution}

\author{Catherine~Lee}
\affiliation{Research Laboratory of Electronics, Massachusetts Institute of Technology, Cambridge, MA 02139, USA}
\affiliation{Lincoln Laboratory, Massachusetts Institute of Technology, Lexington, MA 02420, USA}

\author{Darius~Bunandar}
\affiliation{Research Laboratory of Electronics, Massachusetts Institute of Technology, Cambridge, MA 02139, USA}

\author{Zheshen~Zhang}
\affiliation{Research Laboratory of Electronics, Massachusetts Institute of Technology, Cambridge, MA 02139, USA}

\author{Gregory~R.~Steinbrecher}
\affiliation{Research Laboratory of Electronics, Massachusetts Institute of Technology, Cambridge, MA 02139, USA}
\affiliation{Lincoln Laboratory, Massachusetts Institute of Technology, Lexington, MA 02420, USA}

\author{P.~Ben~Dixon}
\affiliation{Lincoln Laboratory, Massachusetts Institute of Technology, Lexington, MA 02420, USA}

\author{Franco~N.~C.~Wong}
\affiliation{Research Laboratory of Electronics, Massachusetts Institute of Technology, Cambridge, MA 02139, USA}

\author{Jeffrey~H.~Shapiro}
\affiliation{Research Laboratory of Electronics, Massachusetts Institute of Technology, Cambridge, MA 02139, USA}

\author{Scott~A.~Hamilton}
\affiliation{Lincoln Laboratory, Massachusetts Institute of Technology, Lexington, MA 02420, USA}

\author{Dirk~Englund}
\affiliation{Research Laboratory of Electronics, Massachusetts Institute of Technology, Cambridge, MA 02139, USA}

\date{\today}

\maketitle

Quantum key distribution (QKD) exploits the quantum nature of light to share provably secure keys, allowing secure communication in the presence of an eavesdropper. The first QKD schemes used photons encoded in two states, such as polarization \cite{BB84,JCryptology.5.3}. Recently, much effort has turned to large-alphabet QKD schemes, which encode photons in high-dimensional basis states \cite{PRA.61.062308,SRep.3.2316,PRL.84.4737,PRL.98.060503,OptExp.21.15959,2014.PRA.Lee.do-qkdPulsed,2015.NewJPhys.Zhong.hd-qkd,PRA.88.032305,NewJPhys.17.033033}. Compared to binary-encoded QKD, large-alphabet schemes can encode more secure information per detected photon, boosting secure communication rates, and also provide increased resilience to noise and loss \cite{PRL.88.127902}. High-dimensional encoding may also improve the efficiency of other quantum information processing tasks, such as performing Bell tests \cite{NPhys.7.677} and implementing quantum gates \cite{NPhys.5.134}. Here, we demonstrate a large-alphabet QKD protocol based on high-dimensional temporal encoding. We achieve record secret-key rates and perform the first field demonstration of large-alphabet QKD. This demonstrates a new, practical way to optimize secret-key rates and marks an important step towards transmission of high-dimensional quantum states in deployed networks.

\begin{figure*}
\begin{center}
  \includegraphics[scale=0.675]{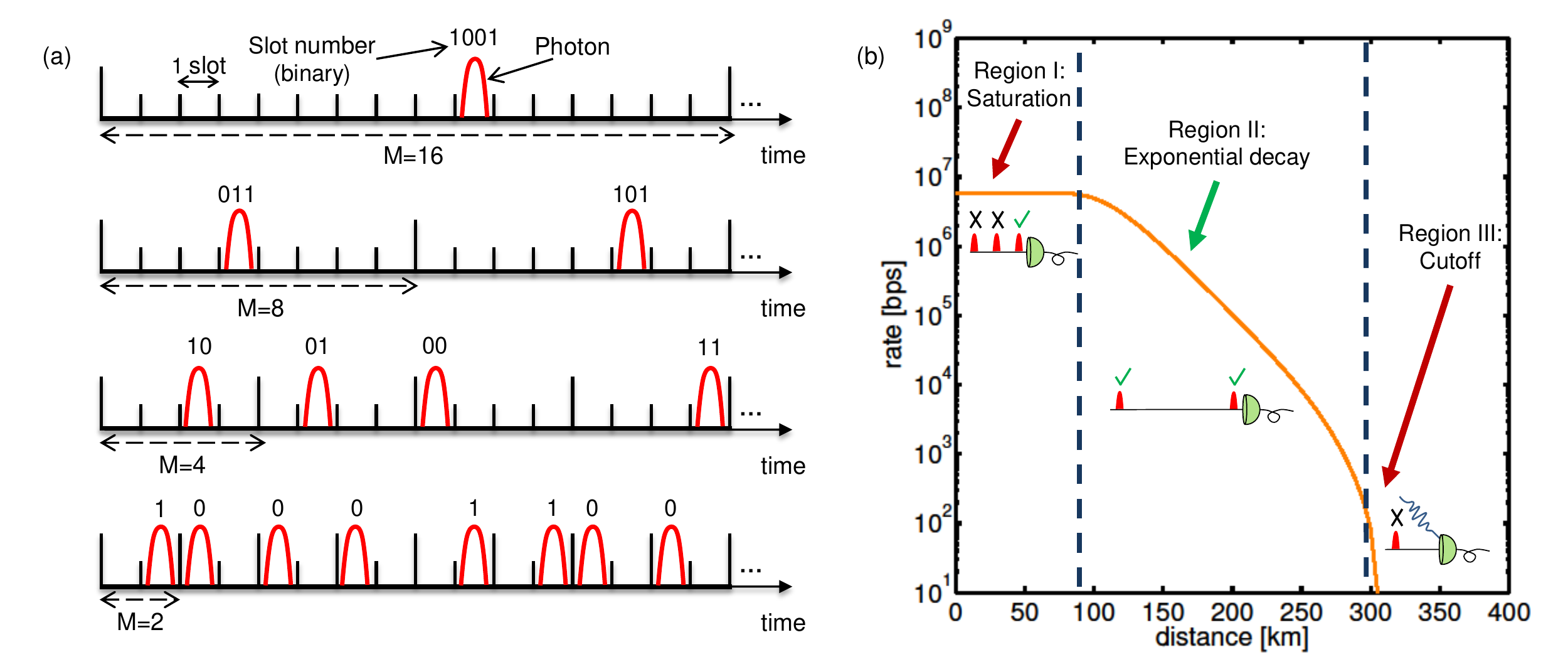}
    \caption{(a) 
    In high-dimensional temporal encoding (pulse position modulation), information is encoded in the position of an optical pulse within $M$ slots, depicted here for alphabet size $M \in \{2,4,8,16\}$. For a fixed slot duration, the alphabet size and the transmitted pulse rate are inversely proportional. 
    (b) Representative plot of secret-key rate versus channel length for a traditional two-dimensional QKD protocol, assuming a 5~Gbps modulation rate, a 0.2~dB/km channel loss, a 1~kcps background count rate, a 93\% detector efficiency, and a 100~ns detector reset time after each detection event. 
    Three regions are denoted: 
    I. At short distances, 0-100~km (or correspondingly, low losses, 0-20~dB), the secret-key rate is limited by detector saturation. 
    II. For higher losses (normal operation), the secret-key rate decays exponentially with distance. 
    III. At even higher losses ($>$ 300~km), a cutoff is reached when Bob's received photon rate becomes comparable to his detectors' background count rate. The error rate grows and the secret-key rate drops abruptly. 
    }
    \label{whyHD}
\end{center}
\end{figure*}

High-dimensional encoding is possible in a variety of degrees of freedom, 
and large-alphabet QKD has been demonstrated in the laboratory using 
position-momentum \cite{SRep.3.2316}, 
time-energy \cite{PRL.84.4737,PRL.98.060503,OptExp.21.15959,2014.PRA.Lee.do-qkdPulsed,2015.NewJPhys.Zhong.hd-qkd}, 
and orbital angular momentum modes \cite{PRA.88.032305,NewJPhys.17.033033}. 
Of these, time-energy encoding is appealing for its compatibility with existing telecommunications infrastructure --- which lowers the barriers to widespread adoption of QKD. 
The time-energy correlations are robust over transmission in both fiber and free-space channels and are preserved when passing through wavelength-division multiplexing. 
In high-dimensional temporal encoding, the position of a photon within a symbol frame comprising $M$ time slots can convey as much as $\log_2 M$ bits of information, as depicted in Figure~\ref{whyHD}(a). 
Classically, this encoding is known as pulse position modulation (PPM), and combined with single-photon detection, it achieves near-optimal performance in terms of bits per detected photon \cite{OptLett.31.444}. 
Assuming a constant slot duration, PPM exhibits a trade-off between the alphabet size $M$ and the transmitted symbol rate: an increase in the former directly corresponds to a decrease in the latter. 
The alphabet size determines how much information is encoded in each photon, and the transmitted symbol rate directly affects how many photons are received per second. 
We take advantage of this trade-off to maximize the secret-key rate in the presence of detector saturation.

Figure~\ref{whyHD}(b) is a representative plot of secret-key rate versus channel length for binary encoding with realizable parameters. 
Three regimes of distance/loss are indicated. 
In normal operation (Region~II), the secret-key rate decreases exponentially with distance until the received photon flux is comparable to the background counts of the detector(s). 
At distances/losses beyond this cutoff point (Region~III), the correlations between sender and receiver are masked by the background and the secret-key rate drops abruptly. 
However, at short distances, i.e., low losses (Region~I), the secret-key rate is limited when the received photon flux saturates the detectors, as illustrated in Figure~\ref{whyHD}(b). 
In this regime, which extends to approximately 100~km for these parameters, the best strategy to maximize the secret-key rate is to reduce the transmitted photon rate by increasing the alphabet size until the detectors are just below saturation. 
Although much research has focused on extending the range of QKD links well beyond 100~km  \cite{NewJPhys.11.075003,OptLett.37.1008,NPhot.9.163}, 
deployed QKD networks will include a variety of link lengths with potentially different optimal technologies,
and thus we focus here on using high-dimensional encoding to maximize secret-key rates over metropolitan-area distances of tens of kilometers.

To demonstrate high-rate, large-alphabet QKD, 
we implemented dispersive-optics QKD (DO-QKD) \cite{2013.PRA.Mower.do-qkd}, a high-dimensional QKD protocol based on time-energy encoding, with the basis transformations produced by group velocity dispersion (GVD).
We previously proved the security of this scheme against arbitrary collective attacks \cite{2013.PRA.Mower.do-qkd} and implemented the scheme using entangled photon pairs in the laboratory \cite{2014.PRA.Lee.do-qkdPulsed}. 
The present work is a prepare-and-measure (P\&M) version of DO-QKD, with decoy-state protection against photon number splitting attacks \cite{PRL.94.230503,PRL.94.230504,PRA.91.022336}.

\begin{figure*}
\begin{center}
    \includegraphics[scale=0.55]{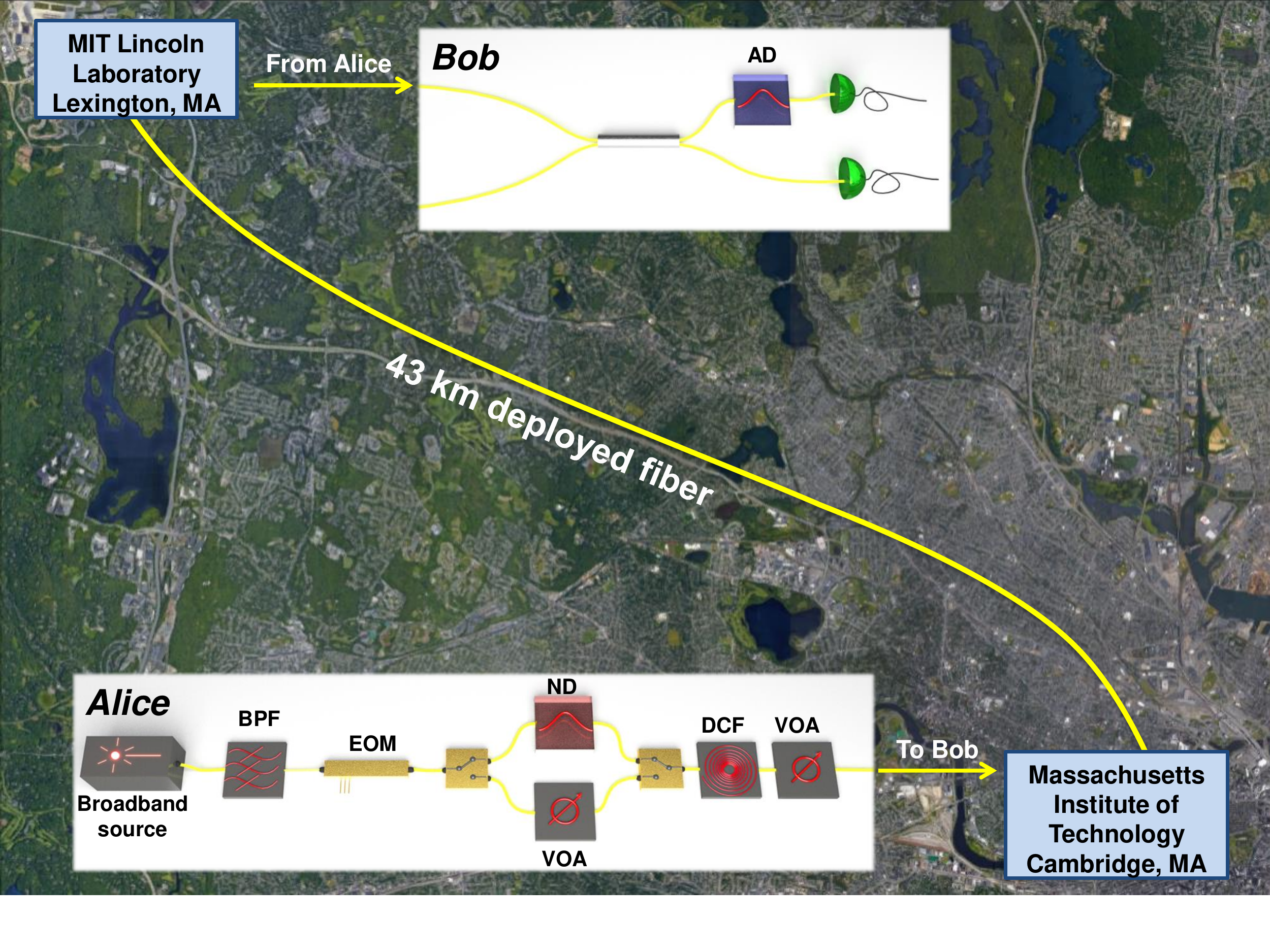}
    \caption{Map showing node locations and approximate path of the installed 43-km deployed-fiber testbed used in this work. Overlaid are Alice's transmitter, located in Cambridge, MA, and Bob's receiver, located in Lexington, MA. 
    BPF: bandpass filter. EOM: electro-optic modulator. VOA: variable attenuator. ND: normal GVD. AD: anomalous GVD. DCF: dispersion-compensating fiber.
    }
    \label{setup}
\end{center}
\end{figure*}

In P\&M DO-QKD, as pictured in Figure~\ref{setup}, the transmitter, Alice, filters a broadband light source to $\sim25$ GHz centered around 1559~nm and uses an electro-optic modulator to encode a PPM sequence that will become the raw key. 
To prepare in the time basis, Alice sends the PPM pulse to the receiver, Bob, and to prepare in the energy basis, she applies normal GVD with magnitude 10,000~ps/nm to the pulse before sending it to Bob. The basis choice must be random to an eavesdropper, Eve, but known to Alice. 
Before transmitting, Alice attenuates the pulses to keep the average number of photons less than one per pulse, but she varies the intensity between signal states, which are used for generating secure keys, and weaker decoy states, which are used for channel monitoring to guard against a photon-number-splitting attack. Alice also precompensates for the GVD incurred over the fiber channel, or the security of the protocol would be degraded.
On a separate channel (not pictured in Figure~\ref{setup}), Alice sends a periodic strong optical pulse that Bob detects with a photodiode and uses as a timing reference. 
To measure in the time basis, Bob detects the photon arrival time, and to measure in the energy basis, he applies anomalous GVD with magnitude 10,000~ps/nm to the photon before detecting the arrival time. 
Bob's single-photon detectors are niobium nitride (NbN) SNSPDs capable of counting at hundreds of Mcps rates, with 68\% detection efficiency, timing resolution of tens of picoseconds, and few kcps dark count rates \cite{OptExp.21.1440}. A single optical fiber is coupled to four interleaved nanowires, which are read out by a commercial time-to-digital converter (Picoquant Hydraharp 400) with a 80~ns dead time per channel. 
Information can be shared when Alice and Bob both apply GVD or both do not apply GVD. When only one party applies GVD, the correlation between prepared pulse time and measured pulse time is degraded from tens of picoseconds (limited by the detector timing resolution) to nanoseconds (determined by the optical bandwidth and the magnitude of the GVD). 
Alice and Bob convert the photon timing correlations into shared secret keys through a series of classical postprocessing steps. Bob demodulates the PPM signal, and Alice and Bob sift their data to postselect symbols encoded and decoded using the same basis. They correct errors between their symbol strings using a multi-layer low-density parity-check (LDPC) code \cite{2013.ITA.Zhou.ecc}, and they perform privacy amplification to eliminate Eve's information about their shared error-free symbol strings.

The secure photon information efficiency (PIE) quantifies Alice and Bob's information advantage over Eve, who can mount arbitrary collective attacks. 
By measuring the covariance matrix associated with the correlation between prepared pulse time and measured pulse time \cite{2013.PRA.Mower.do-qkd,PRL.112.120506} and by monitoring the fraction of detection events originating from single-photon emission with weak-intensity decoy states \cite{PRL.94.230503,PRL.94.230504,PRA.91.022336}, Alice and Bob can bound the information accessible to Eve. Any information that Alice and Bob share in excess of this bound will be secure, except with a finite failure probability that corresponds to the predetermined security parameter $\varepsilon_s$ \cite{PRL.100.200501,PRA.81.062343,2015.QINP.Lee.finite,JPhysA.49.205301}.

We tested the system, varying the PPM alphabet size $M \in \{4,8,16,32\}$, in three scenarios: in the laboratory in the back-to-back configuration with negligible channel loss, in the laboratory using a 41-km spool of standard single-mode fiber, and in a field test over a 43-km deployed fiber. The deployed-fiber testbed comprised a pair of dark fibers running between the main campus of MIT in Cambridge, MA, and MIT Lincoln Laboratory in Lexington, MA, as illustrated in Figure~\ref{setup}. 
Installed fibers are subject to environmental perturbations, such as temperature fluctuations, that are not present in the laboratory, as well as higher losses due to greater numbers of splices and bends. 
The 41-km fiber spool had a total loss of 7.6 dB, but the loss over the deployed fiber was 12.7~dB --- equivalent to 63.5~km of standard single-mode fiber on a spool (assuming standard loss of 0.2 dB/km).

\begin{table*}
\begin{tabular}{|l|r|r|r|}
\hline
& Back-to-back & 41-km spool & 43-km deployed fiber \\ \hline
Loss (dB) & 0.1 & 7.6 & 12.7 \\ 
Slot duration (ps) & 240 & 240 & 240 \\ 
Optimal $M$ & 16 & 8 & 4 \\ 
Max. secret-key rate (bps) & $23 \times 10^6$ & $5.3 \times 10^6$ & $1.2 \times 10^6$ \\
Secure PIE (bit/photon) & 1.40 & 0.88 & 0.50 \\ 
\hline
\end{tabular}
\caption{Summary of the maximum secret-key rates obtained in the three test cases.}
\label{compare}
\end{table*}

\begin{figure}
\begin{center}
     \includegraphics[scale=0.33]{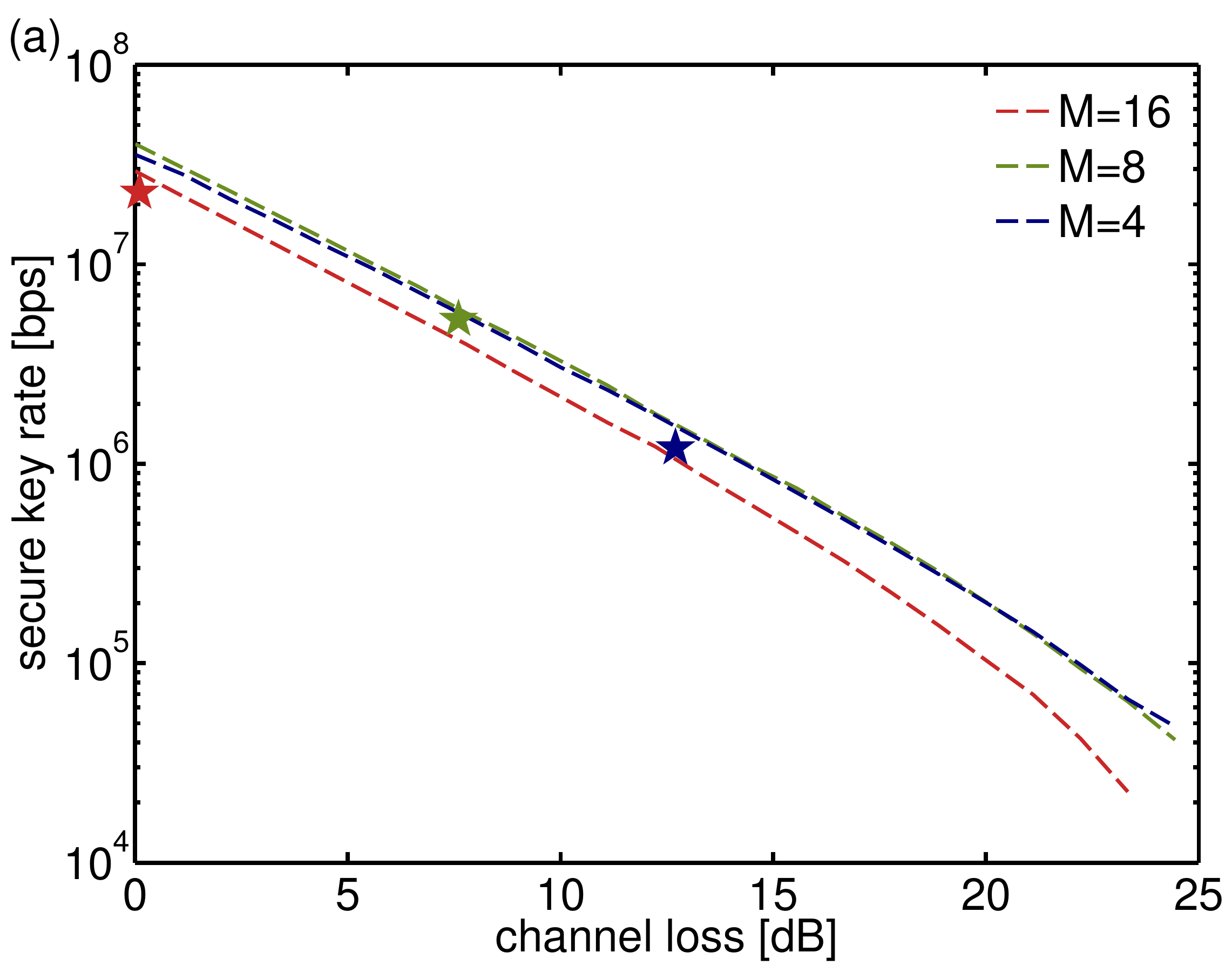}
     \includegraphics[scale=0.33]{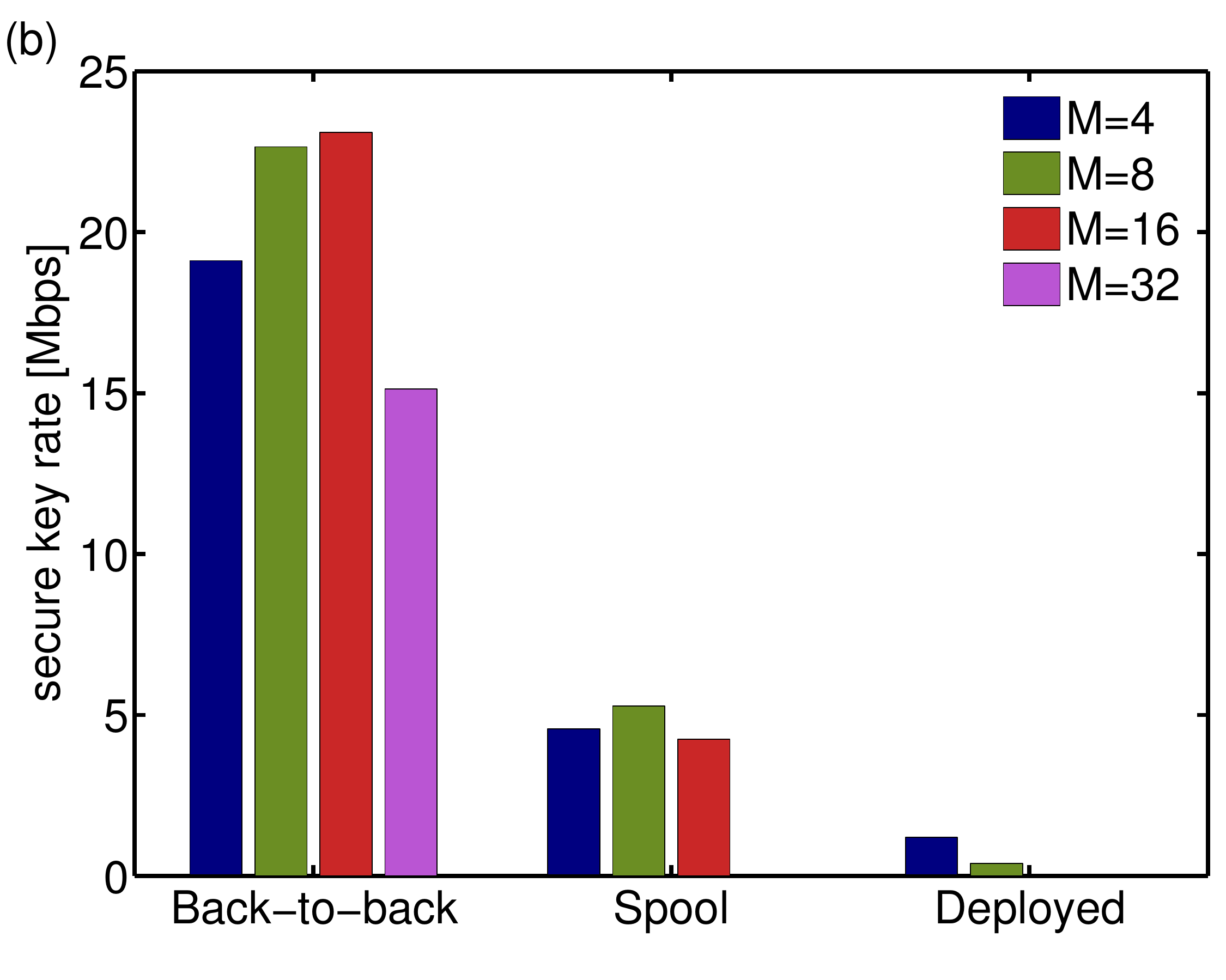}
    \caption{(a) Experimental (stars) and theoretical (dashed curves) secret-key rates versus channel loss. Colors correspond to optimal alphabet size $M$ for each of the three test configurations. Theoretical rates used as inputs the experimental parameters of each of the test configurations. 
     (b) Experimental secret-key rates for all alphabet sizes of each test case. Loss increases from left to right. The optimal $M$ decreases as loss increases. For experimental convenience, we did not increase the alphabet size once it became apparent that doing so would not increase the secret-key rate.
     }
    \label{sbps}
\end{center}
\end{figure}

In the back-to-back configuration, we observed a maximum secret-key rate of 23~Mbps with $M=16$. Over the 41-km fiber spool, the maximum secret-key rate was 5.3~Mbps with $M=8$. 
Over the 43-km deployed fiber, the maximum secret-key rate was 1.2~Mbps with $M=4$. 
Table~\ref{compare} summarizes the three test cases, and 
Figure~\ref{sbps}(a) plots the experimental results along with theoretical secret-key rates as functions of channel loss. 
The reported values and theoretical curves include decoy state and finite-key analysis with sample size $N =10^9$ counts and security parameter $\varepsilon_s = 10^{-10}$ \cite{2015.QINP.Lee.finite,JPhysA.49.205301}. 
Colors correspond to alphabet size and thus to test configuration, since each configuration had a different optimal alphabet size. 
The theoretical curves were computed using the experimental conditions, such as detector timing jitter and the measured timing correlations, which were not the same for all three test configurations. 
Thus, we cannot directly compare the three curves to determine the universally optimal alphabet size for a given loss. 
Instead, Figure~\ref{sbps}(b) displays the secret-key rates obtained for each alphabet size in the three test cases. 

The optimal $M$ to maximize the secret-key rate depends most strongly on Bob's received photon rate, which is in turn a function of channel loss. If Bob had ideal detectors, the highest secret-key rate would be obtained for the fastest transmitter rate, which occurs for $M = 2$. With finite detector reset times, Bob's receivable photon rate is limited, and in the case of detector saturation, increasing $M>2$ allows Alice and Bob to effectively produce secret keys even during the reset time, which can be as long as tens or hundreds of nanoseconds. 
Thus, at short distances and correspondingly low losses, we can expect a bottleneck due to the maximum count rate of Bob's detectors. 
In this detector-limited regime, it is advantageous to increase $M$ to encode as much information as possible in each detected photon while keeping Bob's detectors just below saturation, 
and indeed, Figure~\ref{sbps}(b) demonstrates that the optimal $M$ decreases as channel loss increases.

\begin{figure}
\begin{center}
      \includegraphics[scale=0.33]{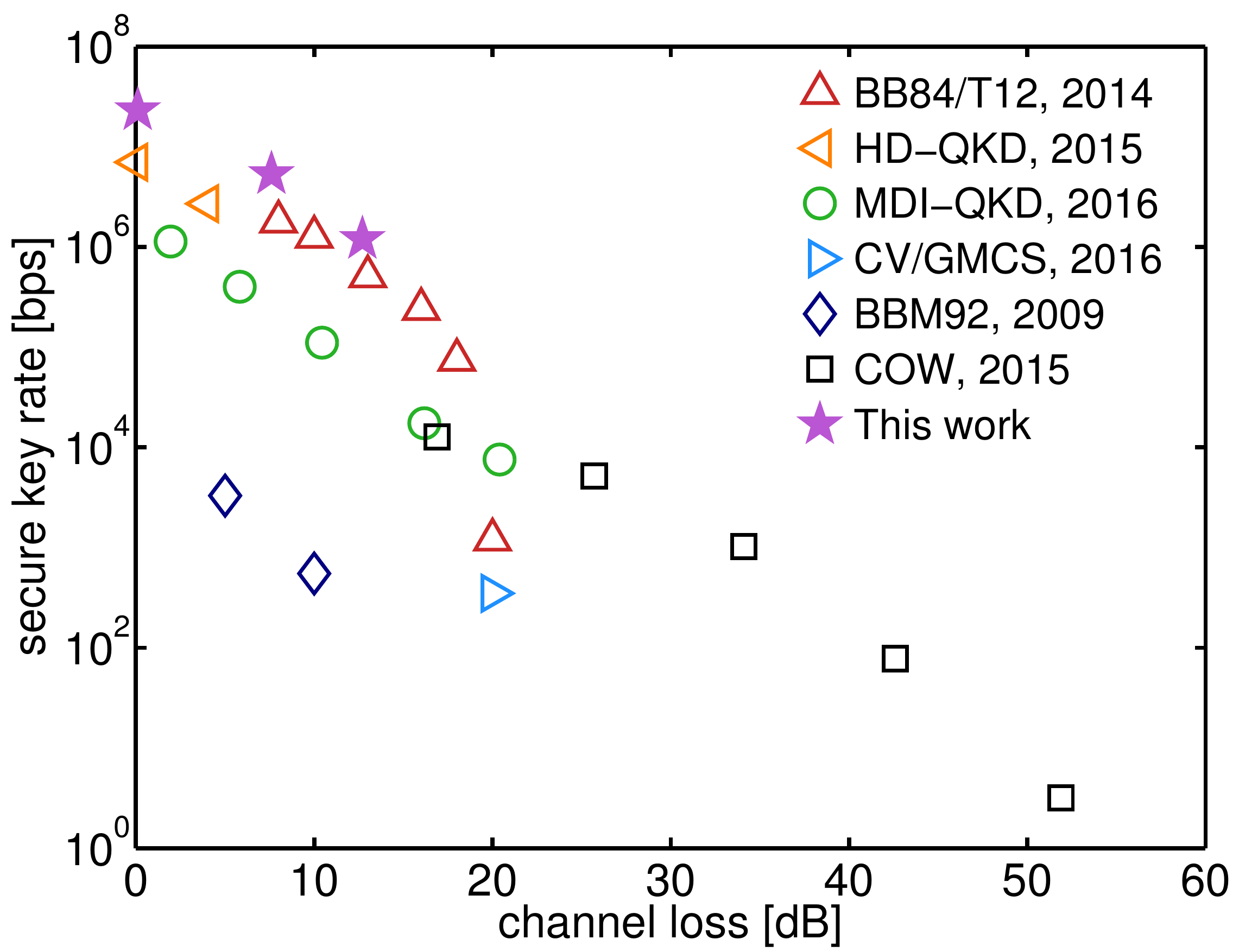}
     \caption{Comparison of our P\&M DO-QKD results to previously published QKD system records, chosen to represent either secure throughput or distance records for a variety of protocols. 
     BB84/T12: secure throughput record for two-dimensional QKD \cite{APL.104.021101}. HD-QKD: secure throughput record for high-dimensional entanglement-based QKD \cite{2015.NewJPhys.Zhong.hd-qkd}. MDI-QKD: secure throughput record for measurement-device-independent QKD \cite{NPhot.10.312}. CV/GMCS: distance record for continuous-variable QKD \cite{SciRep.6.19201}. BBM92: secure throughput record for two-dimensional entanglement-based QKD \cite{NewJPhys.11.045013}. COW: distance record for QKD \cite{NPhot.9.163}.
    }
    \label{records}
\end{center}
\end{figure}

The 1.2~Mbps secret-key rate over the deployed fiber is the highest rate reported to date in a QKD field test 
and also compares favorably to previously published high-rate laboratory demonstrations under similar losses \cite{OptExp.21.24550,APL.104.021101}. 
Additionally, Figure~\ref{records} plots our results along with a variety of notable QKD demonstrations \cite{APL.104.021101,2015.NewJPhys.Zhong.hd-qkd,NPhot.10.312,SciRep.6.19201,NewJPhys.11.045013,NPhot.9.163}. 
Our results show an improvement over other works for channel losses in the range of 0-15~dB. 
Our secret-key rate advantage comes from both the high-dimensional QKD protocol, which effectively generates secure information even during the single-photon detectors' dead time, taking advantage of what would be wasted time for traditional two-dimensional protocols, 
and the fast SNSPDs, which are capable of counting up to hundreds of Mcps \cite{OptExp.21.1440}. 
Slower detectors with longer dead times would amplify the inherent advantage of the high-dimensional protocol, as the detectors would saturate at lower incoming photon rates.

The high-dimensional time-energy encoding demonstrated here offers the ability to optimize the secret-key rate by varying the alphabet size $M$ in response to both channel loss and receiver limitations. 
This is particularly useful when Bob's detectors are saturated, which often occurs over metropolitan-area distances of tens of kilometers. 
By presenting and demonstrating a new protocol intended to adapt to the constraints of a particular link implementation, 
this work represents a new approach to high-rate secure quantum communication optimized for use in metropolitan areas.

\section*{Methods}

\subsection*{Experimental setup}
The deployed-fiber testbed comprised a pair of dark fibers, one of which is used for quantum signals, and the other of which is used for bright synchronization pulses.
Alice's light source was a superluminescent diode with tens of nanometers of optical bandwidth. 
This source can enable DO-QKD with multiple spectral channels, although this demonstration used only one channel with 25~GHz of optical bandwidth, filtered by a tunable bandpass filter. 
The 25~GHz output was modulated by an electro-optic modulator with a PPM sequence of 50~ps pulses centered in 240~ps time slots that was produced by a pulse pattern generator (PPG). The resulting optical pulses were attenuated to either $\mu = 0.5$ photons/pulse for signal states or $\nu = 0.05$ photons/pulse for decoy states. 
A circulator at the output of Alice's transmitter (not pictured in Fig.~\ref{setup}) provided some protection against a Trojan horse attack.
The bright synchronization pulse was produced by a continuous-wave laser modulated by an electro-optic modulator driven by another output of the same PPG. The synchronization pulse period was a constant multiple of the symbol frame length. In the back-to-back and spool tests, the period was 256 times the symbol frame length for all $M$. For the deployed-fiber test, the period was reduced to 64 times the symbol frame length to mitigate the effects of timing drifts over the installed fiber.

Because only one SNSPD system was available, Bob could not randomly choose between the two measurement bases. Therefore, we fixed both Alice and Bob's basis selections for the duration of each data acquisition period. 
The resulting datasets were combined in postprocessing. 
For each test case, numerical optimization of the secret-key rate determined the probabilities with which Alice and Bob should have selected each basis; the data from different bases were combined using these probabilities to compute the reported experimental secret-key rates.
Similarly, Alice's choice of signal or decoy intensity was fixed for the duration of each acquisition period, 
the probabilities with which Alice selected signal or decoy states were determined by numerical optimization for each test case, and the data from different intensities were combined using these probabilities in postprocessing.

\subsection*{Secure photon information efficiency}
In the asymptotic regime, the secure PIE with decoy-state analysis is 
\begin{equation}
 r_{\infty,\mathrm{decoy}} = \beta I(A;B) - (1-F_\mu^\mathrm{LB})\log_2 M - F_\mu^\mathrm{LB}\chi^\mathrm{UB}(A;E),
\end{equation}
where $F_\mu^\mathrm{LB}$ is a lower bound on the fraction of Bob's detection events that came from a single-photon transmission by Alice and $\chi^\mathrm{UB}(A;E)$ is an upper bound on Eve's Holevo information \cite{2013.PRA.Mower.do-qkd,PRL.112.120506,PRA.91.022336}. Decoy state measurements contribute to the estimation of $F_\mu^\mathrm{LB}$ and $\chi^\mathrm{UB}(A;E)$. 
In the finite-key regime, we must consider the effects of a finite sample size on the estimation of the parameters related to decoy states \cite{JPhysA.49.205301}, 
in addition to the standard finite-size effects on parameter estimation, error correction, and privacy amplification \cite{2015.QINP.Lee.finite}.

\section*{Acknowledgements}

This material is based upon work supported by the Office of the Assistant Secretary of Defense for Research and Engineering under Air Force Contract No. FA8721-05-C-0002 and/or FA8702-15-D-0001. 
Any opinions, findings, conclusions or recommendations expressed in this material are those of the author(s) and do not necessarily reflect the views of the Assistant Secretary of Defense for Research and Engineering. 
D.E. and D.B. acknowledge partial support from the Air Force Office of Scientific Research Multidisciplinary University Research Initiative (FA9550-14-1-0052) and the Air Force Research Laboratory RITA program (FA8750-14-2-0120). 
D.B. also acknowledges support from the Samsung Advanced Institute of Technology. 
C.L. thanks David O. Caplan and Nivedita Chandrasekaran for helpful discussions.


%

\end{document}